\documentclass[%
 amsmath,amssymb,
 aps, physrev,
]{revtex4-2}

\usepackage{graphicx}
\usepackage{dcolumn}
\usepackage{bm}
\usepackage{aas_macros}
\usepackage{subcaption}
\usepackage{placeins}
\usepackage{float}
\usepackage{ulem}
\usepackage{caption}
\usepackage{ragged2e}

\usepackage{color}

\begin{document}

\preprint{APS/123-QED}

\title{\textbf{Turbulent Transport of Galactic Magnetic Fields into Cosmic Voids: \\ Insights from IllustrisTNG}}%

\author{Yuri Yamashita$^1$}
\email{Contact author: yamashita.yuri.h4@s.mail.nagoya-u.ac.jp}
\author{Hiroyuki Tashiro$^1$}
\author{Kiyotomo Ichiki$^{1,2,3}$}
\affiliation{
$^1$Graduate School of Science, Nagoya University, Furocho, Chikusa-ku, Nagoya, Aichi, 464-8602, Japan\\
$^2$Kobayashi-Maskawa Institute for the Origin of Particles and the Universe, Nagoya University, Furocho, Chikusa-ku, Nagoya, Aichi, 464-8602, Japan\\
$^3$Institute for Advanced Research, Nagoya University, Furocho, Chikusa-ku, Nagoya, Aichi 464-8602, Japan
}

\date{\today}

\begin{abstract}
Astrophysical processes associated with galaxies may contribute to the magnetization of cosmic voids. We investigate the diffusion of galactic magnetic fields in an expanding universe in the presence of turbulent magnetic diffusivity. To estimate the turbulent magnetic diffusivity, we analyze the TNG50-1 data set of the IllustrisTNG simulation and derive its redshift dependence from the characteristic turbulent velocity and turbulent scale of the intergalactic medium.
Using the resulting diffusivity, we find a present-day magnetic screening length of $\sim 7\,{\rm Mpc}$, 
roughly twenty times larger than previous estimates based on a constant turbulent diffusivity due to the larger turbulent scale obtained in our analysis. This scale corresponds to a significant fraction of the characteristic size of cosmic voids and suggests that galactic magnetic fields can play a more important role in void magnetization than previously estimated.
\end{abstract}

\maketitle

\section{Introduction}
Magnetic fields are observed throughout the Universe across various spatial scales. However, the origin of the magnetic fields that are believed to exist in cosmic voids remains unclear ~\cite{neronov2010evidence,dolag2010lower,dermer2011time,takahashi2012lower,vovk2012fermi,tashiro2013cosmological,acciari2023lower}. 
While the upper limit on the magnetic field in cosmic voids is constrained to be about $1\,\mathrm{nG}$ by the cosmic microwave background observations~\cite{minoda2021small,2016A&A...594A..19P} and the large-scale structure of the universe~\cite{2025PhRvL.135g1001P}, the lower limit has been estimated to be
$\gtrsim 10^{-17}\,\mathrm{G}$ for a coherence length $\lambda_{B}\gtrsim0.2~\mathrm{Mpc}$ based on observations of secondary cascades from TeV gamma rays from blazars ~\cite{2010Sci...328...73N,2012ApJ...744L...7T, 2013ApJ...771L..42T,2023A&A...670A.145A} and from a gamma-ray burst~\cite{2026PhRvD.113d3041B}.
We note that gamma-ray lower limits on magnetic fields in voids are model-dependent, because they rely on assumptions about cascade emission, source activity time, and possible plasma-instability cooling of the pairs \cite{2025arXiv250917104D,2026arXiv260408375D}. Nevertheless, they provide one of the most sensitive probes of weak magnetic fields in cosmic voids.

Several recent studies have investigated whether galactic magnetic fields can contribute to, or even account for, the magnetic fields observed in cosmic voids.
For example, Ref.~\cite{garg2025magnetic} argued that the dipole components of galactic magnetic fields can extend deep into cosmic voids and may be sufficiently strong to satisfy the observational lower limit on void magnetic fields.
In contrast, Ref.~\cite{seller2025contribution} concluded that such contamination is insufficient to explain the inferred void magnetic fields once plasma screening effects are taken into account. In their analysis, the magnetic diffusivity was assumed to be determined solely by the Spitzer resistivity associated with the finite conductivity of the Universe.

More recently, attention has turned to the role of turbulence in enhancing magnetic diffusion. Turbulent magnetic diffusion represents the effective transport of magnetic fields by random motions in a conducting plasma and can greatly exceed the diffusion associated with the Spitzer resistivity. Such turbulence may be generated by nonlinear structure formation as well as astrophysical feedback processes, including galactic outflows and AGN activity. Ghosh et al.~\cite{ghosh2025can} incorporated turbulent magnetic diffusivity and found that galactic magnetic fields can diffuse over scales of order $\sim 0.1\,\mathrm{Mpc}$, substantially larger than estimates based solely on the Spitzer resistivity. Nevertheless, they concluded that such diffusion remains insufficient to explain the required volume filling factor of magnetic fields in cosmic voids.

The turbulent magnetic diffusivity depends sensitively on the characteristic turbulent velocity and turbulent scale of the intergalactic medium. Since these quantities are governed by nonlinear structure formation and astrophysical feedback processes, their amplitudes remain highly uncertain. In previous studies, the turbulent diffusivity was estimated using simplified assumptions.

Motivated by this uncertainty, we revisit the diffusion of galactic magnetic fields into cosmic voids using a turbulent magnetic diffusivity estimated directly from cosmological simulations. In this work, we estimate the turbulent velocity and turbulent scale of the intergalactic medium using the TNG50-1 data~\cite{2019MNRAS.490.3196P,2019MNRAS.490.3234N} from the IllustrisTNG simulation~\cite{pillepich2018first,pillepich2018simulating,marinacci2018first,nelson2018first} and use them to evaluate the redshift evolution of the turbulent magnetic diffusivity. Combining this estimate with an analytical solution of the induction equation, we investigate the propagation of galactic magnetic fields into cosmic voids.

This paper is organized as follows. In Sec. II, we present the theoretical framework for magnetic-field diffusion in an expanding universe. We derive an analytic solution to the diffusion equation for an initially localized galactic magnetic field. In Sec. III, we introduce our estimate of the turbulent magnetic diffusivity from the TNG50-1 data in the IllustrisTNG simulation. In Sec. IV, we discuss the physical implications of the resulting diffusion scales, compare them with previous estimates, and comment on the possible magnetic contamination of cosmic voids. Our conclusions are summarized in Sec. V.
Throughout this paper, we adopt the same $\Lambda$CDM cosmology as used in the IllustrisTNG simulations:
the total matter density  $\Omega_{\rm m}=0.31$,
the dark energy density $\Omega_\Lambda=0.69$,
and
the Hubble constant $H_0=100\,h\,{\rm km\,s^{-1}\,Mpc^{-1}}$
with the dimensionless Hubble constant $h=0.67$.

\section{Diffusion of Galactic Magnetic Fields into Voids}

The main goal of this paper is to investigate whether galactic magnetic fields can be transported from galaxies into cosmic voids through magnetic diffusion in a turbulent plasma. As a first step, we derive an analytical solution of the induction equation in an expanding universe.

To describe the diffusion of large-scale magnetic fields from galaxies into cosmic voids, we consider the induction equation in Fourier space in an expanding universe, including the effect of turbulent magnetic diffusivity
\cite{ghosh2025can,jedamzik2011evolution}:
\begin{equation}
    \partial_z \bm{B}_{\bm{k}}(z)
    = \frac{2}{1+z}\,\bm{B}_{\bm{k}}(z)
    + [\eta(z)+\eta_\mathrm{turb}(z)]\,
    \frac{1+z}{H(z)}\,k^2\,\bm{B}_{\bm{k}}(z)\,.
    \label{eq:diffusion}
\end{equation}
Here $\eta$ is the magnetic diffusivity due to the finite conductivity of the Universe, $\eta_{\rm turb}$ is the turbulent magnetic diffusivity, $H(z)$ is the Hubble parameter at redshift $z$, and $\bm{B}_{\bm{k}}$ is the Fourier transform of the magnetic field,
\begin{equation}
\bm{B}_{\bm{k}}(z)
=
\int { d}^3 {\bm r}\,
e^{-i\bm{k}\cdot\bm{r}}
\bm{B}(\bm{r},z),
\end{equation}
where $\bm r$ denotes the comoving position vector and $\bm k$ is the comoving wavenumber vector with $k=|\bm k|$.
Throughout this paper, $\bm{B}$, $\eta$, and $\eta_{\rm turb}$ are defined in physical coordinates.

The analytic solution of Eq.~\eqref{eq:diffusion} is given by 
\begin{equation}
    \bm{B}_{\bm{k}}(z)
    = \bm{B}_{\bm{k}}(z_\mathrm{ini})
    \left(\frac{1+z}{1+z_\mathrm{ini}}\right)^{\!2}
    \exp\!\bigg[
    -k^2\int_z^{z_\mathrm{ini}}\!\mathrm{d}\tilde z\,
    \frac{(\eta(\tilde z)+\eta_\mathrm{turb}(\tilde z))\,(1+\tilde z)}
    {H(\tilde z)}
    \bigg]\, ,
    \label{eq:analytical}
\end{equation}
where $z_{\rm ini}$ is the initial redshift.

The magnetic diffusivity $\eta$ is assumed to be given by the Spitzer conductivity and is expressed as
\begin{equation}
\eta \simeq 10^{7}
\left(\frac{T_{\mathrm{b}}}{10^{4}\,\mathrm{K}}\right)^{-3/2}
\left(\frac{\ln \Lambda_{\mathrm{c}}}{30}\right)
\ \mathrm{cm}^{2}\,\mathrm{s}^{-1},
\end{equation}
where $T_b$ denotes the baryon temperature and $\ln \Lambda_{\mathrm{c}}$ is the Coulomb logarithm. 
By substituting the present-day baryon temperature $T_{\mathrm{b}} \simeq 3000\,\mathrm{K}$, Ref.~\cite{ghosh2025can} obtained
$\eta \simeq 10^{-19}\,\mathrm{kpc\,km\,s^{-1}}$.
On the other hand, the typical turbulent magnetic diffusivity was estimated to be
$\eta_{\mathrm{turb}} \simeq 10^{4}\,\mathrm{kpc\,km\,s^{-1}}$
in Ref.~\cite{ghosh2025can}. Therefore, the contribution of $\eta$ is entirely negligible compared with $\eta_{\rm turb}$. In the following, we neglect $\eta$ and retain only the turbulent magnetic diffusivity.
We will later demonstrate, using IllustrisTNG, that the assumption $\eta \ll \eta_{\mathrm{turb}}$ is indeed valid.

Applying the inverse Fourier transform to Eq.~\eqref{eq:analytical},
\begin{equation}
\bm{B}(\bm{r},z)=\frac{1}{(2\pi)^3}
\int {d}^3 {\bm k}\, e^{i\bm{k}\cdot\bm{r}}
\bm{B}_{\bm{k}}(z),
\end{equation}
we obtain
\begin{equation}
\label{eq:B_screened_conv}
    \bm{B}(\bm{r}, z)
    = \frac{k_\mathrm{s}^3}{(4\pi)^{3/2}}
    \left(\frac{1+z}{1+z_\mathrm{ini}}\right)^{\!2}
    \int_{-\infty}^{\infty}\!{d}^3 {\bm r}'\;
    \bm{B}(\bm{r}',z_\mathrm{ini})
    \exp\!\bigg[
    -\frac{k_\mathrm{s}^2}{4}
    \left|\bm{r}-\bm{r}'\right|^2
    \bigg]\, .
\end{equation}
Here, we introduce the screening wavenumber
$k_\mathrm{s}$ defined through
\begin{equation}
    k_\mathrm{s}(z)
    = \bigg[
    \int_z^{z_\mathrm{ini}}\!{d}\tilde z\,
    \frac{\eta_\mathrm{turb}(\tilde z)\,(1+\tilde z)}
    {H(\tilde z)}
    \bigg]^{-1/2}\, ,
\label{eq:screening_wavenumber}
\end{equation}
and the magnetic screening length is then defined as $l_{\rm s}=2\pi/k_{\rm s}$ \cite{jedamzik2011evolution}. 
This represents the characteristic length scale over which magnetic fields diffuse from their source and become exponentially suppressed.

For simplicity, we assume a purely azimuthal magnetic field of constant magnitude within the galaxy as the initial condition,
\begin{equation}
\bm{B}(\bm{r},z_\mathrm{ini})
=\bm{B}_0\,\Theta(r_\mathrm{g}-r),
\end{equation}
where $r=|\bm r|$, $r_{\rm g}$ is the comoving radius of the initially magnetized galactic region, $\Theta$ is the Heaviside step function and
\begin{equation}
\bm{B}_0=B_0\hat{\phi},
\end{equation}
with $B_0$ denoting the magnitude of the initial magnetic field and $\hat{\phi}$ the azimuthal unit vector. This configuration satisfies the divergence-free condition,
$\nabla\cdot\bm{B}=0$.

Substituting this initial condition into Eq.~\eqref{eq:B_screened_conv}, we obtain
\begin{equation}
\label{eq:B_screened_uniform_sphere}
    \bm{B}(\bm{r}, z)
    = \frac{k_\mathrm{s}^3}{(4\pi)^{3/2}}
    \left(\frac{1+z}{1+z_\mathrm{ini}}\right)^{\!2}
    \bm{B}_0
    \int_{-\infty}^{\infty}\!{d}^3 {\bm r}'\;
    \Theta(r_\mathrm{g}-r')\,
    \exp\!\bigg[
    -\frac{k_\mathrm{s}^2}{4}
    \left|\bm{r}-\bm{r}'\right|^2
    \bigg]\, .
\end{equation}

Evaluating the Gaussian integral yields
\begin{equation}
\label{eq:B_screened_integral}
    \bm{B}(\bm{r}, z)
    = \bm{B}_{0}\left(\frac{1+z}{1+z_\mathrm{ini}} \right)^{\!2}
    \frac{k_{\rm s}}{2\sqrt{\pi}\,r}
    \int_{0}^{r_\mathrm{g}}\!{d} r'\;
    r'\left[
    \exp\!\left(-\left(\frac{k_{\rm s}}{2}\right)^2(r-r')^2\right)
    - \exp\!\left(-\left(\frac{k_{\rm s}}{2}\right)^2(r+r')^2\right)
    \right]\, .
\end{equation}
Carrying out the radial integral gives
\begin{multline}
\label{eq:B_screened_final}
\bm{B}(\bm{r},z)
= \bm{B}_{0}\left(\frac{1+z}{1+z_\mathrm{ini}} \right)^{\!2}
\Bigg[
\frac{1}{2}
\left\{
\operatorname{erf}\left[\frac{k_{\rm s}}{2} (r_{\rm g}-r)\right]
+\operatorname{erf}\left[\frac{k_{\rm s}}{2} (r_{\rm g}+r)\right]
\right\} \\
-\frac{1}{\sqrt{\pi}\,r\,k_{\rm s}}
\left\{
\exp\left[-\left(\frac{k_{\rm s}}{2}\right)^2(r-r_{\rm g})^2\right]
-\exp\left[-\left(\frac{k_{\rm s}}{2}\right)^2(r+r_{\rm g})^2\right]
\right\}
\Bigg]\, ,
\end{multline}
where $\operatorname{erf}(x)$ is the Gaussian error function.

Equation~\eqref{eq:B_screened_final} shows that the characteristic scale of magnetic fields diffusing from galaxies is determined by the screening scale $k_{\rm s}$. Through Eq.~\eqref{eq:screening_wavenumber}, this quantity is controlled by the turbulent magnetic diffusivity, $\eta_{\rm turb}(z)$. Therefore, a realistic estimate of $\eta_{\rm turb}(z)$ is essential for predicting the redshift evolution of the magnetic-field distribution. In the next section, we derive $\eta_{\rm turb}(z)$ from the IllustrisTNG simulation.

\begin{figure*}[t!]
\centering

\begin{subfigure}{0.4\textwidth}
    \centering
    \includegraphics[width=\linewidth]{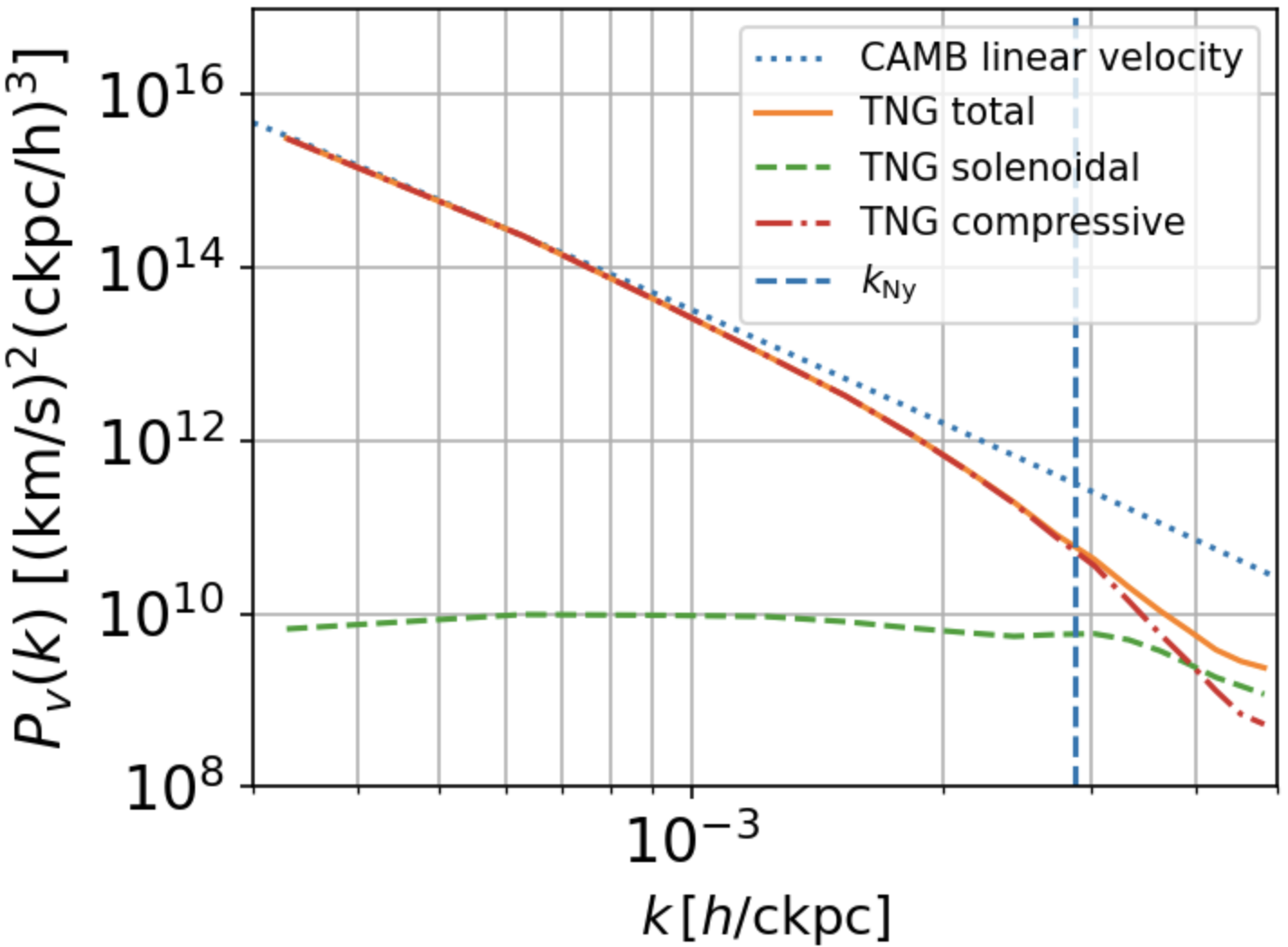}
    \caption{$z=9.39$}
\end{subfigure}
\hfill
\begin{subfigure}{0.4\textwidth}
    \centering
    \includegraphics[width=\linewidth]{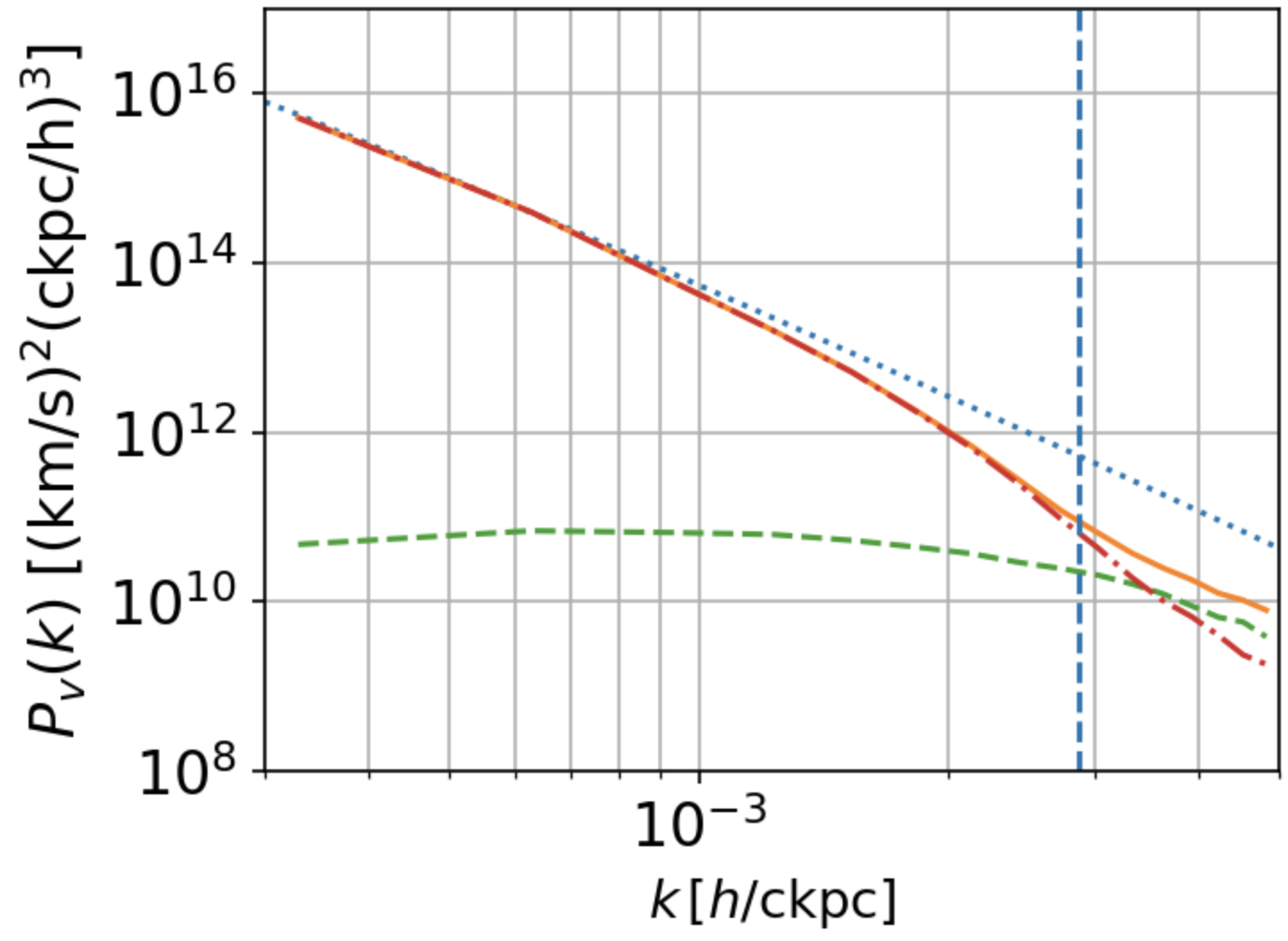}
    \caption{$z=5.23$}
\end{subfigure}

\vspace{0.5cm}

\begin{subfigure}{0.4\textwidth}
    \centering
    \includegraphics[width=\linewidth]{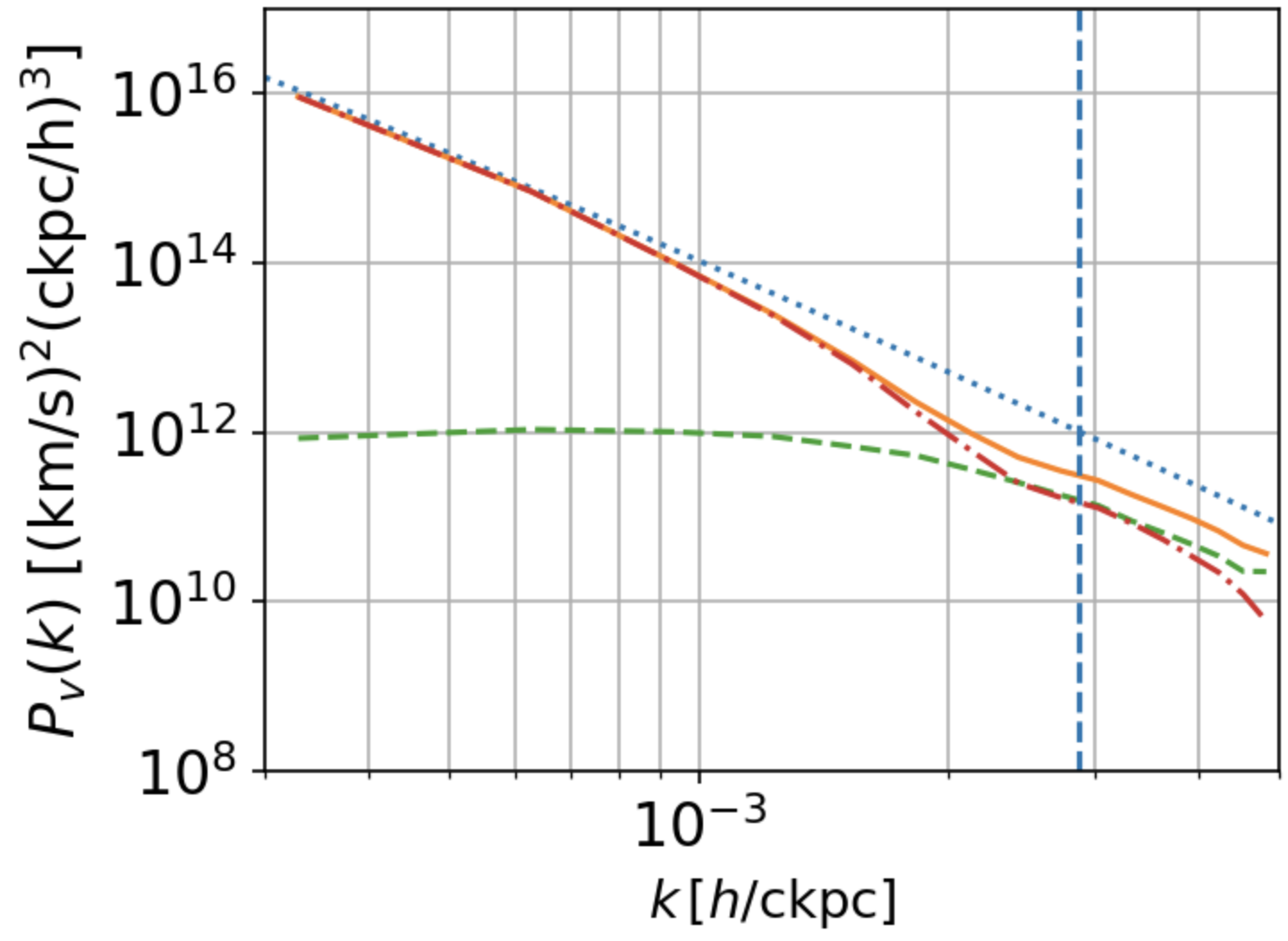}
    \caption{$z=2.10$}
\end{subfigure}
\hfill
\begin{subfigure}{0.4\textwidth}
    \centering
    \includegraphics[width=\linewidth]{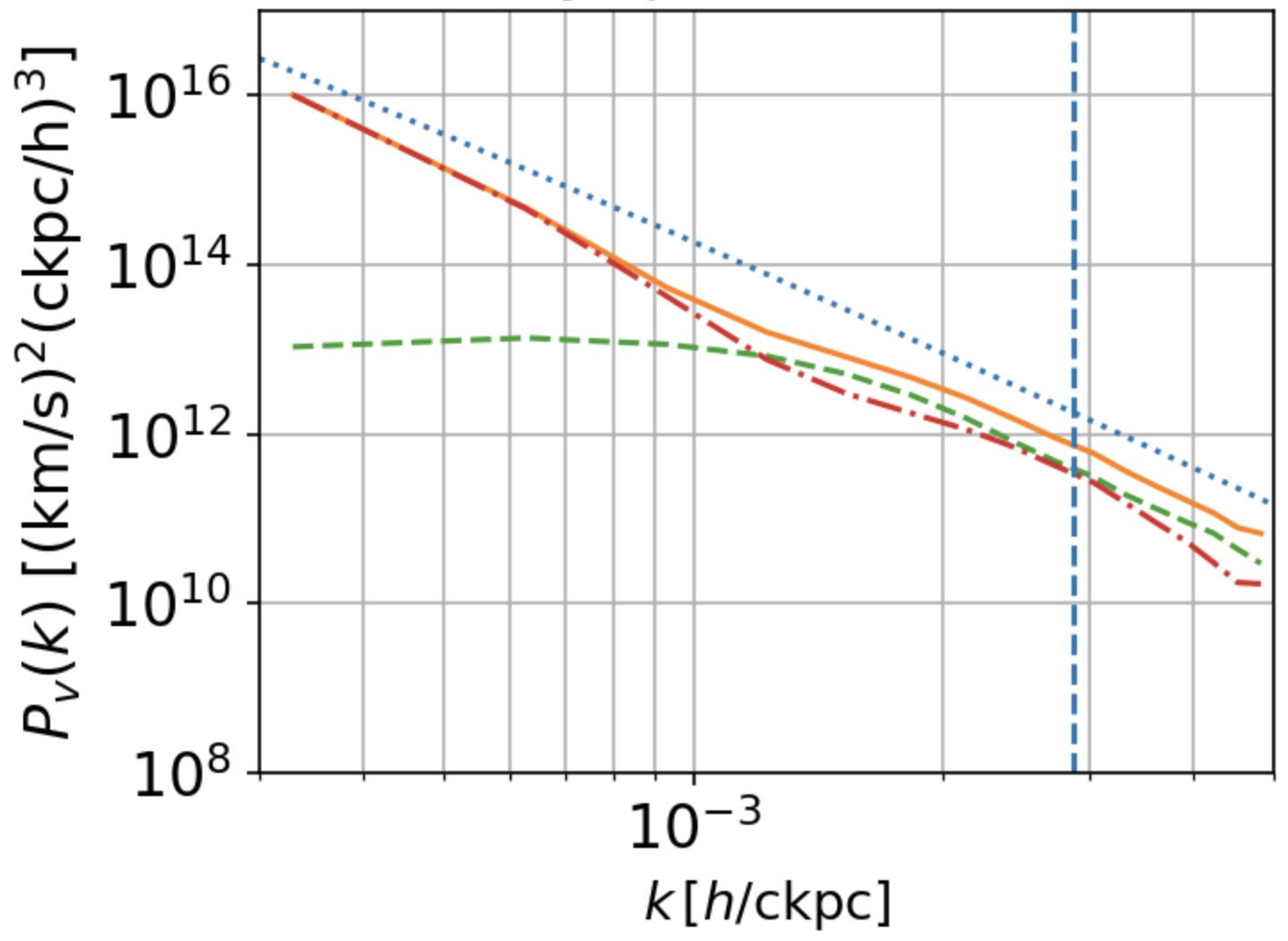}
    \caption{$z=0.01$}
\end{subfigure}

\captionsetup{
    justification=justified,
    singlelinecheck=false
}
\caption{\justifying
Velocity power spectra at different redshifts, obtained from the TNG50-1 data. 
The solid curves show the power spectrum of the total velocity field, $\bm{u}_{\rm tot}$, while the dashed and dash-dotted curves show those of the solenoidal and compressive components, $\bm{u}_{\rm rot}$ and $\bm{u}_{\rm comp}$, respectively. The dotted curve shows the velocity power spectrum predicted by linear theory and calculated using CAMB. The vertical line marks the Nyquist wavenumber, $k_{\rm Ny}=2.87\times10^{-3}\,h\,{\rm ckpc}^{-1}$.
}
\label{fig:u_rot_distribution}
\end{figure*}

\begin{figure*}[t!]
\centering

\begin{subfigure}{0.48\textwidth}
    \centering
    \includegraphics[width=\linewidth]{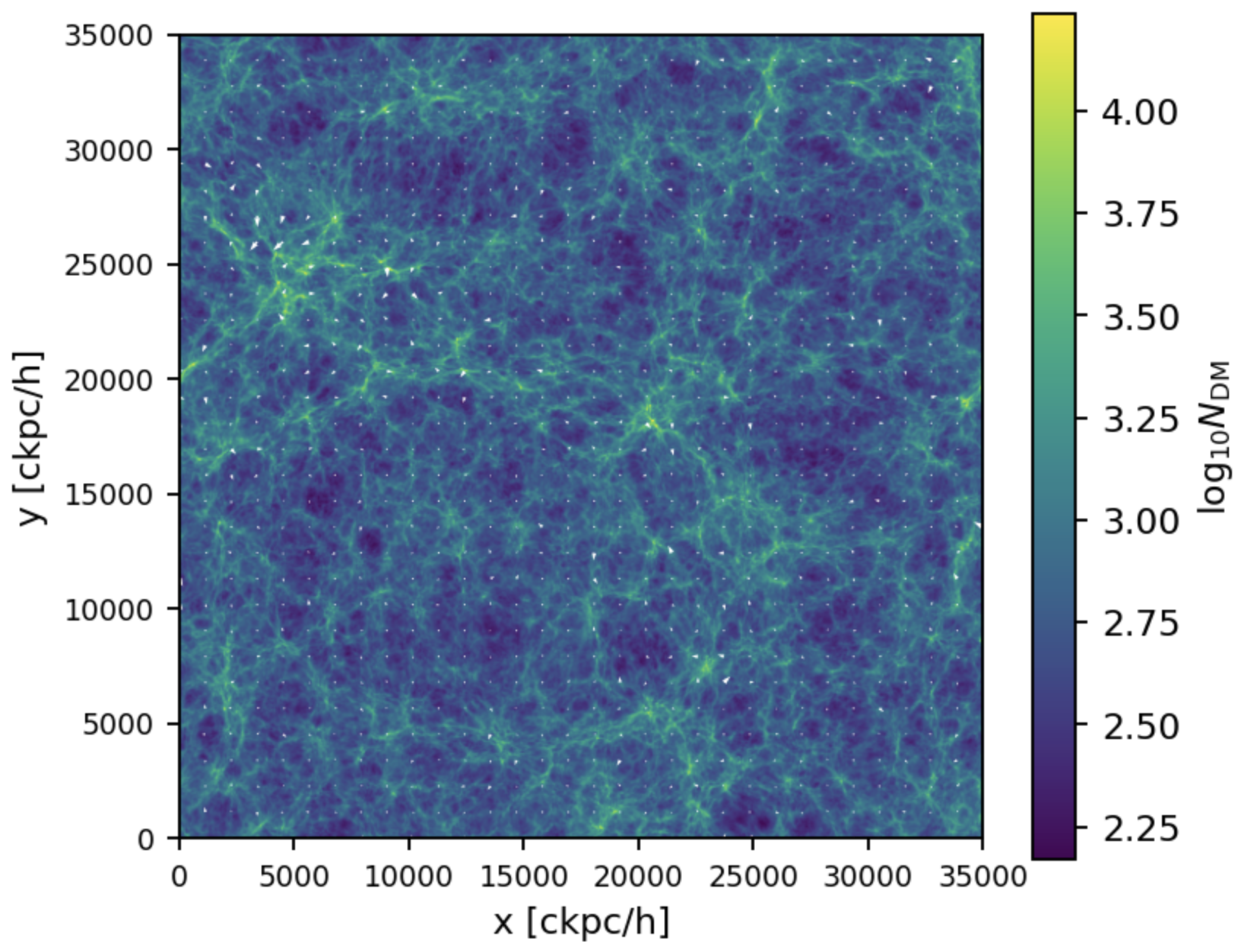}
    \caption{$z=9.39$}
\end{subfigure}
\hfill
\begin{subfigure}{0.48\textwidth}
    \centering
    \includegraphics[width=\linewidth]{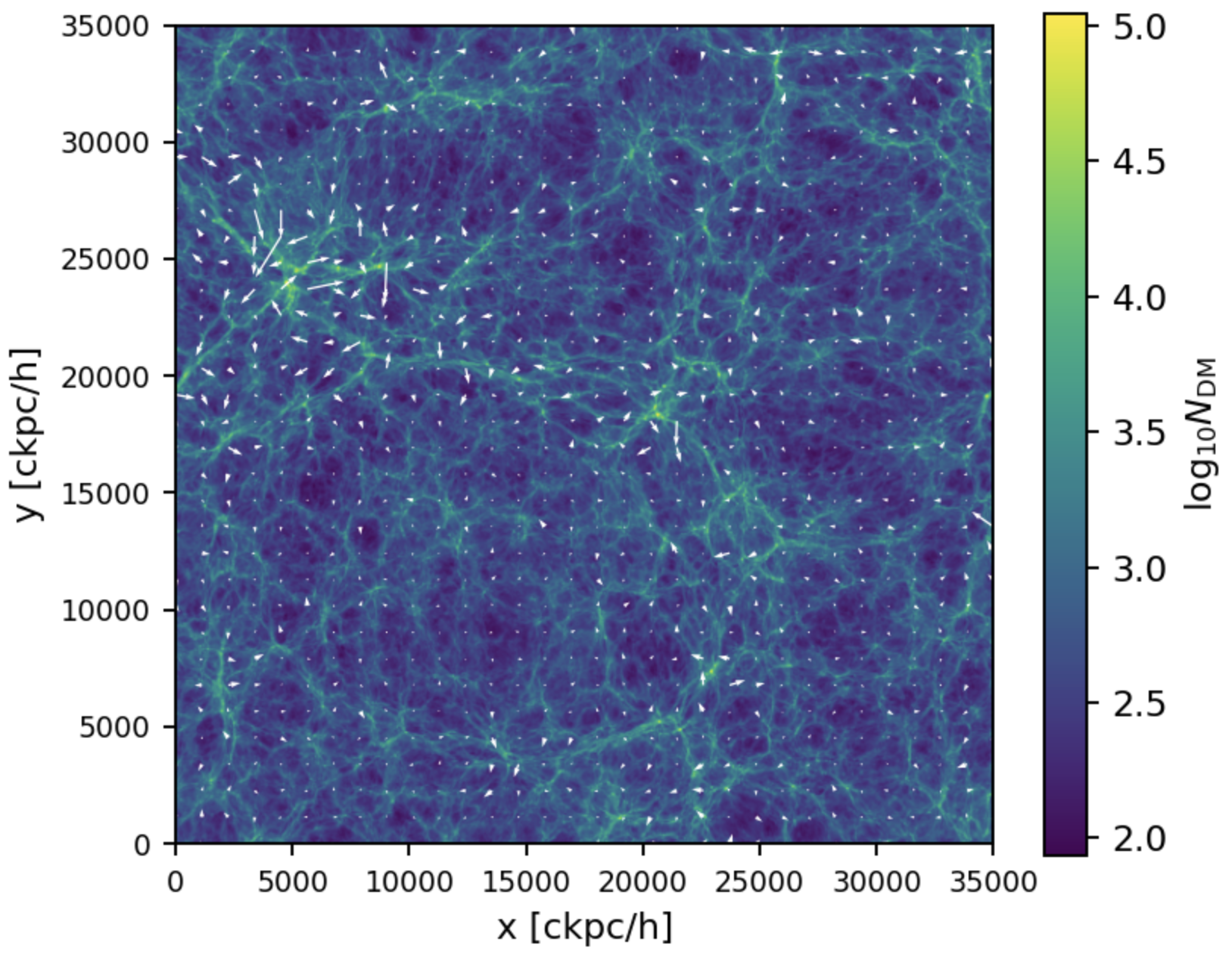}
    \caption{$z=5.23$}
\end{subfigure}

\vspace{0.5cm}

\begin{subfigure}{0.48\textwidth}
    \centering
    \includegraphics[width=\linewidth]{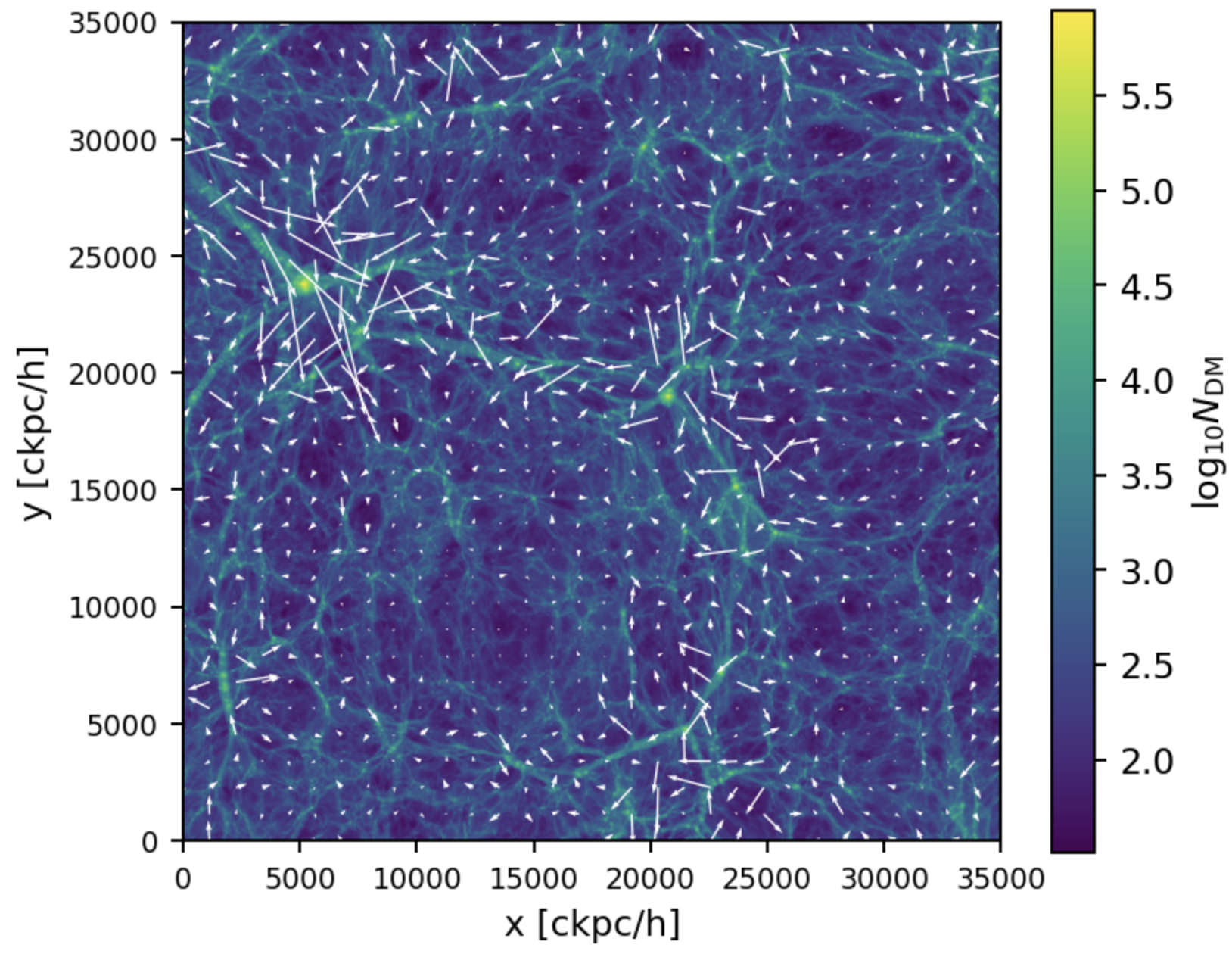}
    \caption{$z=2.10$}
\end{subfigure}
\hfill
\begin{subfigure}{0.48\textwidth}
    \centering
    \includegraphics[width=\linewidth]{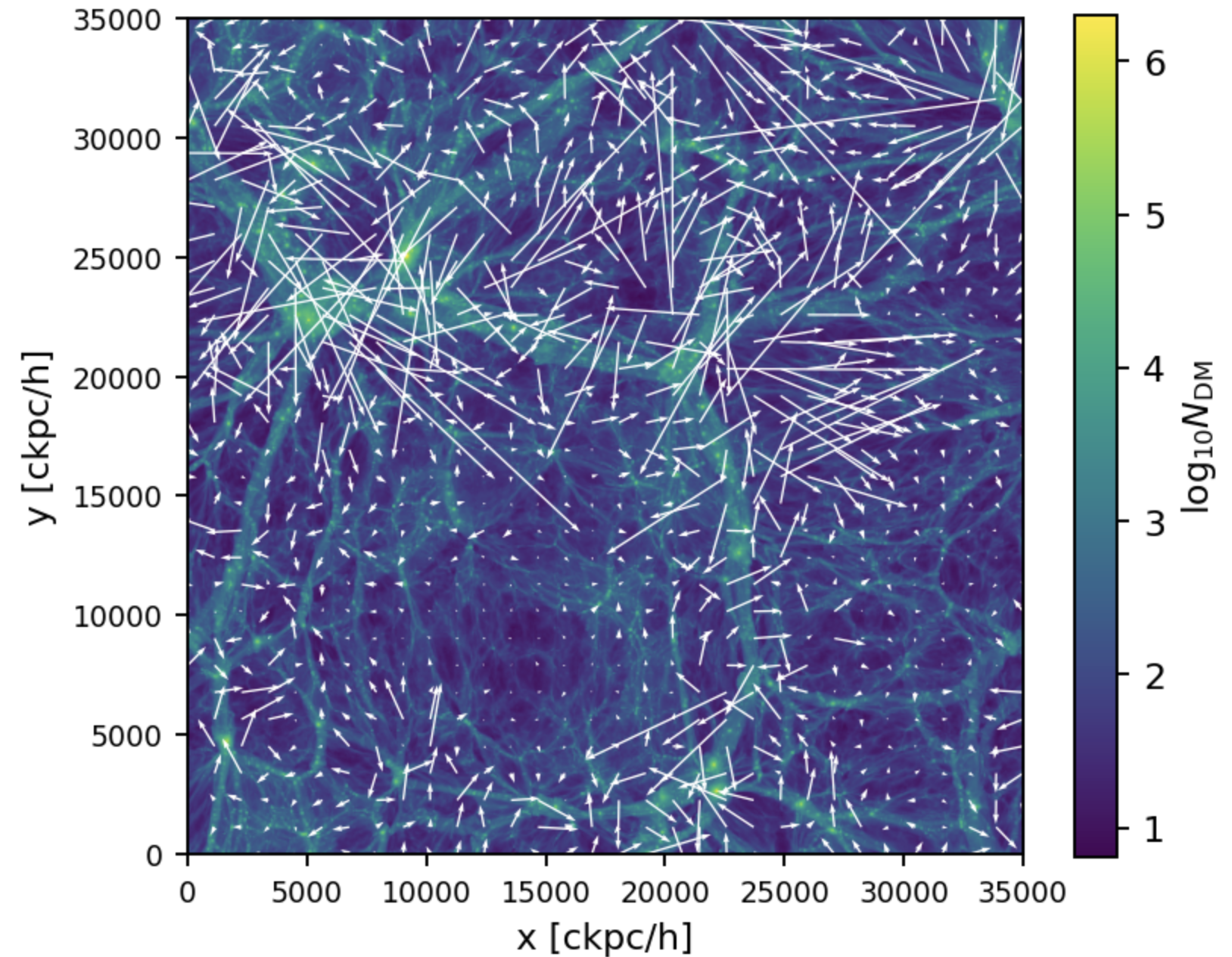}
    \caption{$z=0.01$}
\end{subfigure}

\captionsetup{
    justification=justified,
    singlelinecheck=false
}
\caption{\justifying
Spatial distribution of the turbulent velocity field, $\bm{u}_{\rm rot}$, overlaid on the dark matter distribution at 
different redshifts. The white arrows indicate the in-plane components of the three-dimensional turbulent velocity field. The color map shows $\log_{10}N_{\rm DM}$, where $N_{\rm DM}$ is the projected dark matter particle count assigned to each two-dimensional grid cell using the cloud-in-cell scheme, in a slab centered at $Z=0.5L_{\rm box}$ with a thickness of $0.02L_{\rm box}$.}
\label{fig:u_rot_spatial_distribution}
\end{figure*}

\begin{figure*}[t!]
\centering

\begin{subfigure}{0.48\textwidth}
    \centering
    \includegraphics[width=\linewidth]{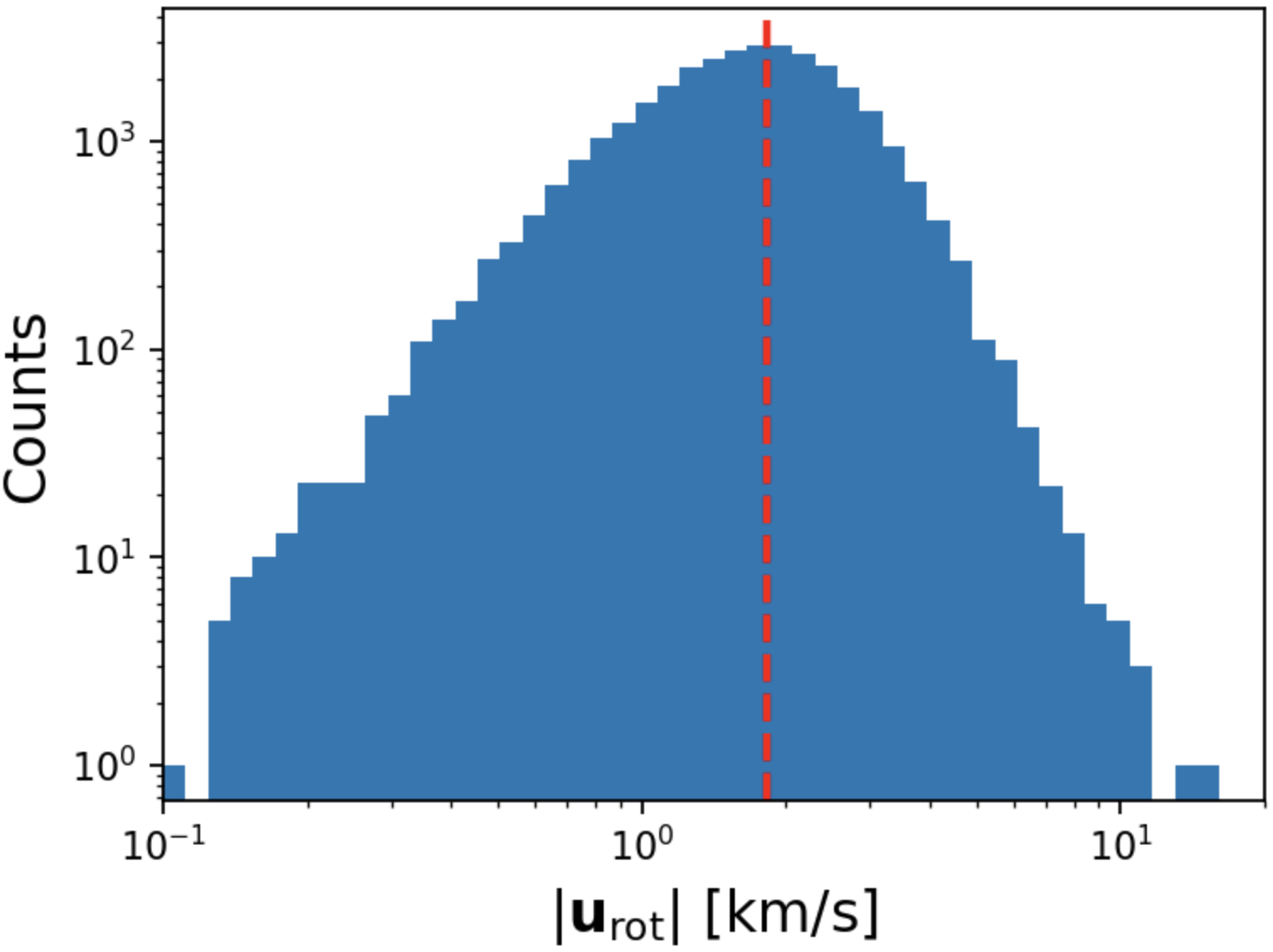}
    \caption{$z=9.39$}
\end{subfigure}
\hfill
\begin{subfigure}{0.48\textwidth}
    \centering
    \includegraphics[width=\linewidth]{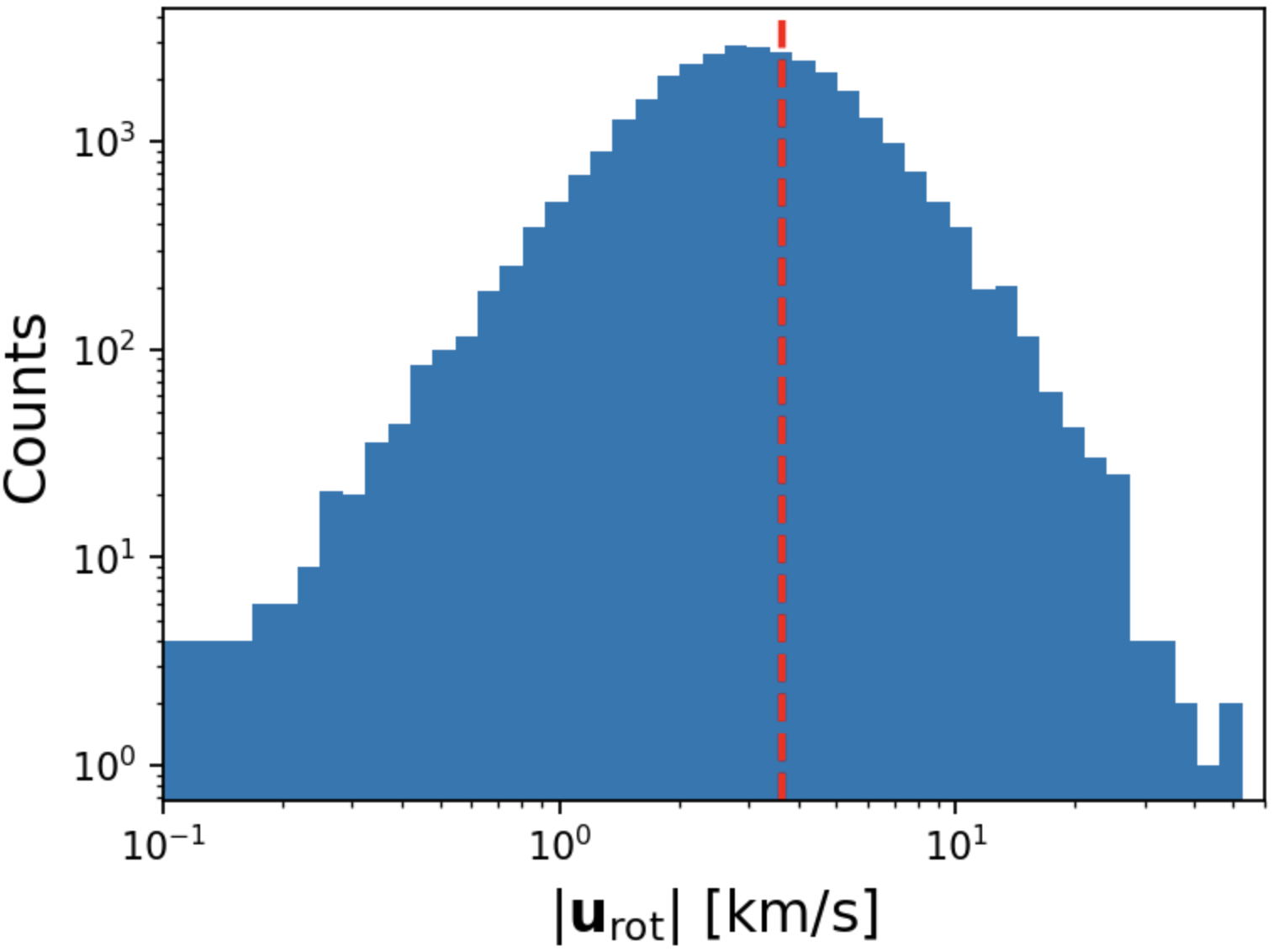}
    \caption{$z=5.23$}
\end{subfigure}

\vspace{0.5cm}

\begin{subfigure}{0.48\textwidth}
    \centering
    \includegraphics[width=\linewidth]{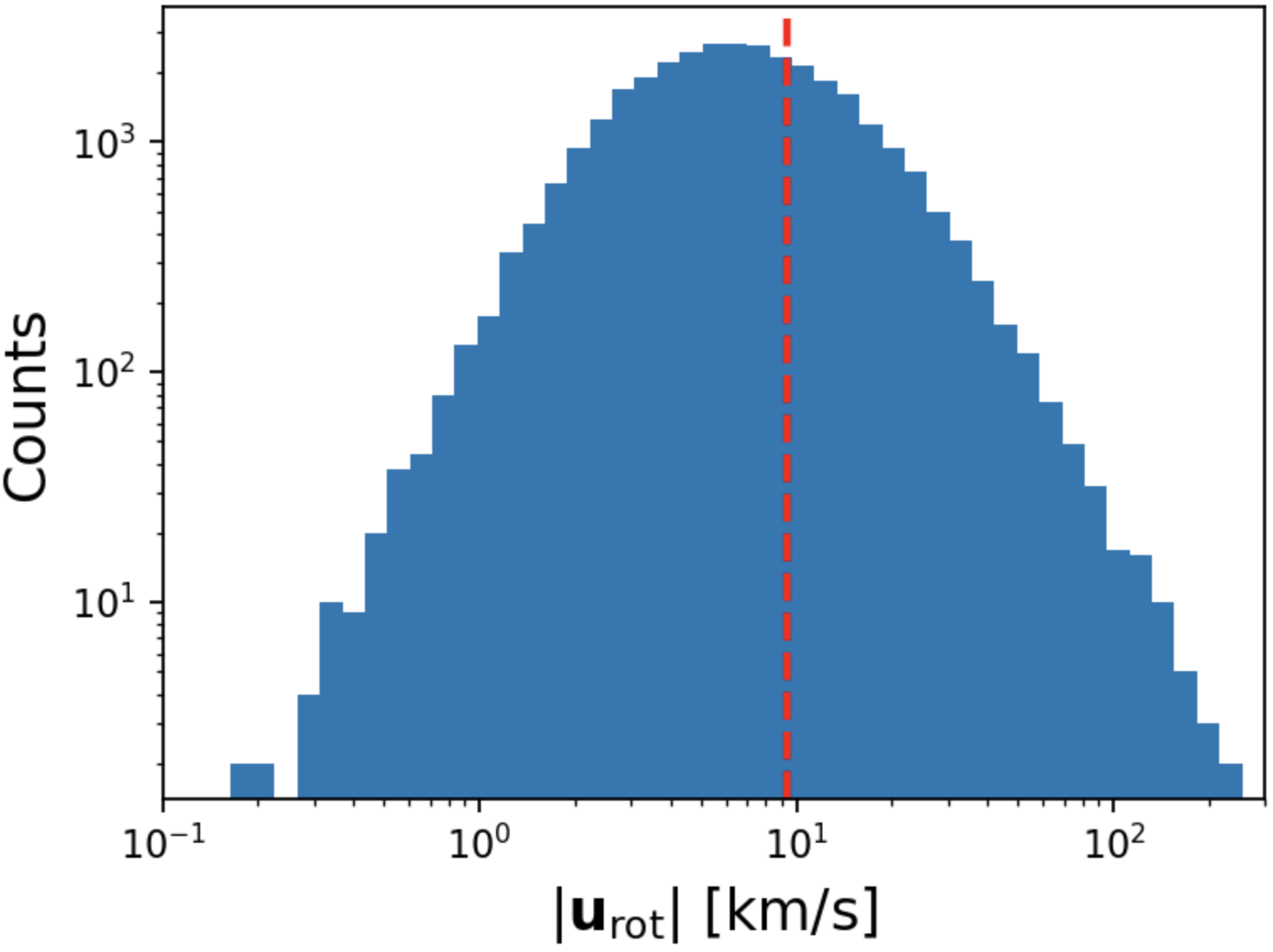}
    \caption{$z=2.10$}
\end{subfigure}
\hfill
\begin{subfigure}{0.48\textwidth}
    \centering
    \includegraphics[width=\linewidth]{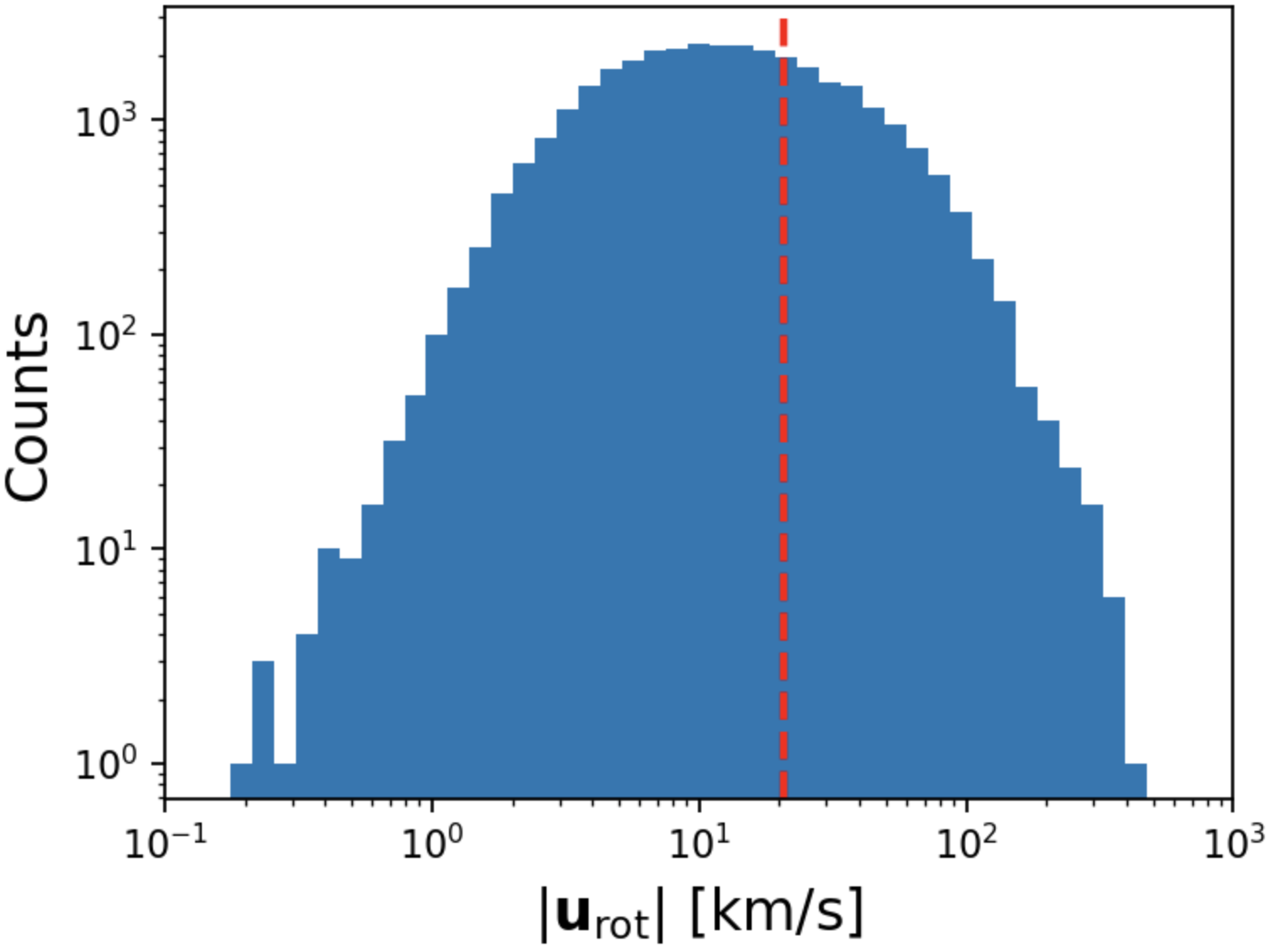}
    \caption{$z=0.01$}
\end{subfigure}

\captionsetup{
    justification=justified,
    singlelinecheck=false
}
\caption{\justifying
Histogram of the magnitude of the rotational velocity field $|\bm{u}_{\rm rot}|$ for baryonic particles at different redshifts. The rotational velocity field is obtained from a Helmholtz decomposition of the mass-weighted velocity field. The red dashed line in each panel denotes the mean value of the distribution, which is adopted as the turbulent velocity $u_{\rm turb}$ in the calculation of the turbulent diffusivity $\eta_{\rm turb}$.
}
\label{fig:u_rot_mean}
\end{figure*}

\section{Estimation of Turbulent Magnetic Diffusivity from TNG50-1}

As shown in Eq.~\eqref{eq:B_screened_final},
the characteristic scale of the magnetic field diffused from a galaxy is determined by $k_{\rm s}$.
To evaluate $k_{\rm s}$ from Eq.~\eqref{eq:screening_wavenumber},
we estimate the redshift-dependent turbulent magnetic diffusivity $\eta_{\rm turb}(z)$ in this section.

\subsection{Identification of Turbulent Motions}

In the previous study~\cite{ghosh2025can},
$\eta_{\rm turb}$ was estimated by assuming characteristic values of the turbulent velocity and length scale.
Here we derive these quantities directly from the state-of-the-art cosmological MHD simulation IllustrisTNG (TNG50-1).

We analyze snapshots at redshifts
$z=9.39$, $5.23$, $2.10$, and $0.01$.
First, in each redshift snapshot,
the gas mass and momentum are assigned to a uniform $32^3$ grid using the cloud-in-cell (CIC) scheme. The CIC scheme distributes the mass and momentum of each gas element among neighboring grid cells according to its position. The mass-weighted velocity in each cell is then obtained by dividing the gridded momentum by the gridded gas mass. The resulting  velocity field is denoted by
$\bm{u}_{\rm tot}$,
and is referred to as the total velocity fluctuation field.

Magnetic diffusion is driven by turbulent motions.
To extract the turbulent component, we 
decompose $\bm{u}_{\rm tot}$
into compressive and solenoidal components, performing a Helmholtz decomposition in Fourier space.
The compressive (longitudinal) component, $\bm{u}_{{\rm comp},\bm{k}}$, and the solenoidal (transverse) component, $\bm{u}_{{\rm rot},\bm{k}}$ are given by
\begin{align}
\bm{u}_{{\rm comp},\bm{k}}
&=
\frac{\bm{k}(\bm{k}\cdot\bm{u}_{{\rm tot},\bm{k}})}{k^2},\\
\bm{u}_{{\rm rot},\bm{k}}
&=
\bm{u}_{{\rm tot},\bm{k}}
-\bm{u}_{{\rm comp},\bm{k}}
=
\bm{u}_{{\rm tot},\bm{k}}
-\frac{\bm{k}(\bm{k}\cdot\bm{u}_{{\rm tot},\bm{k}})}{k^2},
\end{align}
where $\bm{u}_{{\rm tot},\bm{k}}$ is the Fourier components of $\bm{u}_{{\rm tot}}$.

The Helmholtz decomposition used here is closely related to the scalar-vector decomposition in cosmological perturbation theory: the compressive component corresponds to the scalar mode, while the solenoidal component corresponds to the vector mode. Since vector modes are generated through nonlinear structure formation and astrophysical feedback processes, they provide a natural measure of turbulence in the intergalactic medium. Accordingly, we adopt the solenoidal component as the turbulent component throughout this work.

We compute the velocity power spectra of the total, compressive, and solenoidal velocity fields,
\begin{equation}
    \left\langle 
    \bm{u}_{{\rm i}, \bm{k}}
    \cdot
    \bm{u}^* _{{\rm i}, \bm{k'}}
    \right\rangle
    =
    (2\pi)^3
    \delta^{3}
    (\bm{k}-\bm{k}')
    P_{v,{\rm i}} (k),
\end{equation}
where the subscript $\rm i$ denotes the total (${\rm tot}$), compressive (${\rm comp}$), and solenoidal (${\rm rot}$) velocity fields. Note that, in our notation, $P_{v, \rm tot}(k)=P_{v, \rm comp}(k)+P_{v, \rm rot}(k)$.

Figure~\ref{fig:u_rot_distribution} shows the resulting power spectra. The compressive component dominates on large scales and is consistent with the linear-theory prediction obtained from CAMB~\cite{Lewis:1999bs}. In contrast, the solenoidal component becomes increasingly important on small scales.

As the redshift decreases, the amplitude of the solenoidal component grows systematically, indicating the development of turbulent motions during cosmic structure formation. Consequently, the scale at which the solenoidal component overtakes the compressive component shifts toward larger scales at lower redshifts. This suggests that turbulent motions become important over an increasingly broad range of scales as nonlinear structures evolve. Both the scale dependence and the redshift evolution of the solenoidal component support its interpretation as a tracer of turbulent motions generated through nonlinear structure formation and astrophysical feedback processes.

Note that, in Fig.~\ref{fig:u_rot_distribution}, the suppression observed near the Nyquist scale (the vertical dashed line) is likely dominated by finite-resolution effects associated with the CIC assignment and aliasing, rather than by physical damping of velocity fluctuations. Therefore, the estimate of $\eta_{\rm turb}$ in this paper may be conservative, since turbulent motions on unresolved scales are not included.

Applying the inverse Fourier transform to $\bm{u}_{{\rm rot},\bm{k}}$, we reconstruct the solenoidal velocity field, $\bm{u}_{{\rm rot}}$. Since the solenoidal component is adopted as a tracer of turbulent motions throughout this work, $\bm{u}_{\rm rot}$ is regarded as the turbulent velocity field. Figure~\ref{fig:u_rot_spatial_distribution} shows the spatial distribution of  the turbulent velocity field, $\bm{u}_{\rm rot}$, at redshifts $z=9.39$, $5.23$, $2.10$, and $0.01$, overlaid on the dark matter distribution. The white arrows indicate the in-plane components of the three-dimensional turbulent velocity field.

The figure demonstrates that turbulent motions are closely associated with nonlinear structures and are particularly prominent around galaxies and their surrounding environments. As the Universe evolves toward lower redshifts, turbulent flows gradually develop and become stronger. This behavior is consistent with the growth of nonlinear structures through gravitational collapse, mergers, gas accretion, shocks, and astrophysical feedback processes. These results suggest that turbulent motions may contribute to the transport of galactic magnetic fields into surrounding void regions.

\subsection{Evaluation of the Turbulent Magnetic Diffusivity}

When turbulence is approximately isotropic,
the turbulent magnetic diffusivity $\eta_{\rm turb}(z)$ can be estimated as in Ref.~\cite{malkus1979magnetic},
\begin{equation}
\eta_{\rm turb} \simeq \frac{1}{3} u_{\rm turb}\lambda_{\rm turb},
\label{eq:eta_turb_formula}
\end{equation}
where $u_{\rm turb}$ is the characteristic turbulent velocity and
$\lambda_{\rm turb}$ is the characteristic scale of the turbulent motions.

To estimate the turbulent magnetic diffusivity through Eq.~\eqref{eq:eta_turb_formula}, we now evaluate the characteristic turbulent velocity associated with the solenoidal component, $u_{\rm turb}$, from the distribution of $|\bm{u}_{\rm rot}|$ shown in Figure~\ref{fig:u_rot_mean}. The figure shows the histogram of $|\bm{u}_{\rm rot}|$ evaluated over all $32^3$ grid cells after reconstructing the solenoidal velocity field from $\bm{u}_{{\rm rot},\bm{k}}$ at each redshift. 
The mean value of the distribution, indicated by the red vertical line in each panel, is adopted as the characteristic turbulent velocity,
\begin{equation}
u_{\rm turb}
=
\langle |\bm{u}_{\rm rot}| \rangle.
\end{equation}
The corresponding dispersion is estimated as
\begin{equation}
\sigma_u^2 =
\left\langle
\left(
|\bm{u}_{\rm rot}|-u_{\rm turb}
\right)^2
\right\rangle.
\end{equation}

Figure~\ref{fig:u_rot_mean} shows that the distribution of $|\bm{u}_{\rm rot}|$ shifts toward larger velocities with decreasing redshift. This trend is consistent with the redshift evolution of the solenoidal power spectrum shown in Fig.~\ref{fig:u_rot_distribution},
and reflects the growth of nonlinear structures,
where gravitational collapse, gas accretion, mergers, shocks, and astrophysical feedback continuously enhance turbulent motions in the baryonic gas.

Next, we evaluate the characteristic turbulent scale,
$\lambda_{\rm turb}$, from the power spectrum of the solenoidal velocity field,
$P_{v,\rm rot}(k)$.
We define $\lambda_{\rm turb}$ as the characteristic turbulent scale,
\begin{equation}
\lambda_{\rm turb}
=
\frac{
\int dk\,
4\pi k^2 P_{v,\rm rot}(k)\,\lambda(k)
}
{
\int dk\,
4\pi k^2 P_{v,\rm rot}(k)
},
\label{eq:lambda_turb}
\end{equation}
where $\lambda(k)={2\pi}/{k}$.

The corresponding dispersion of the turbulent scale is estimated as
\begin{equation}
\sigma_\lambda^2
=
\frac{
\int dk\,
4\pi k^2 P_{v,\rm rot}(k)
\left[
\lambda(k)-\lambda_{\rm turb}
\right]^2
}
{
\int dk\,
4\pi k^2 P_{v,\rm rot}(k)
}.
\label{eq:sigma_lambda}
\end{equation}
Here, $\sigma_\lambda$ characterizes the width of the turbulent-scale distribution encoded in the power spectrum rather than a statistical measurement uncertainty. The resulting values of $u_{\rm turb}$ and $\lambda_{\rm turb}$ are summarized in Table~\ref{tab:turbulent_diffusivity}. Both quantities increase monotonically toward lower redshifts. This behavior reflects the growth of nonlinear turbulent motions during cosmic structure formation.

Substituting $u_{\rm turb}$ and $\lambda_{\rm turb}$ into Eq.~\eqref{eq:eta_turb_formula}, we obtain the turbulent magnetic diffusivity, whose values are also listed in Table~\ref{tab:turbulent_diffusivity}. Assuming that the turbulent velocity and turbulent scale are uncorrelated, the dispersion of the turbulent magnetic diffusivity is estimated by propagating the dispersions of $u_{\rm turb}$ and $\lambda_{\rm turb}$:
\begin{equation}
\sigma_{\eta}
=
\frac{1}{3}
\sqrt{
\lambda_{\rm turb}^{2}\sigma_{u}^{2}
+
u_{\rm turb}^{2}\sigma_{\lambda}^{2}
}.
\end{equation}

The turbulent magnetic diffusivity increases by more than two orders of magnitude from $z=9.39$ to $z=0.01$
This indicates that turbulent magnetic transport becomes more efficient as nonlinear structures develop during cosmic structure formation.

\begin{figure*}[t!]
\centering

\includegraphics[width=.55\textwidth]{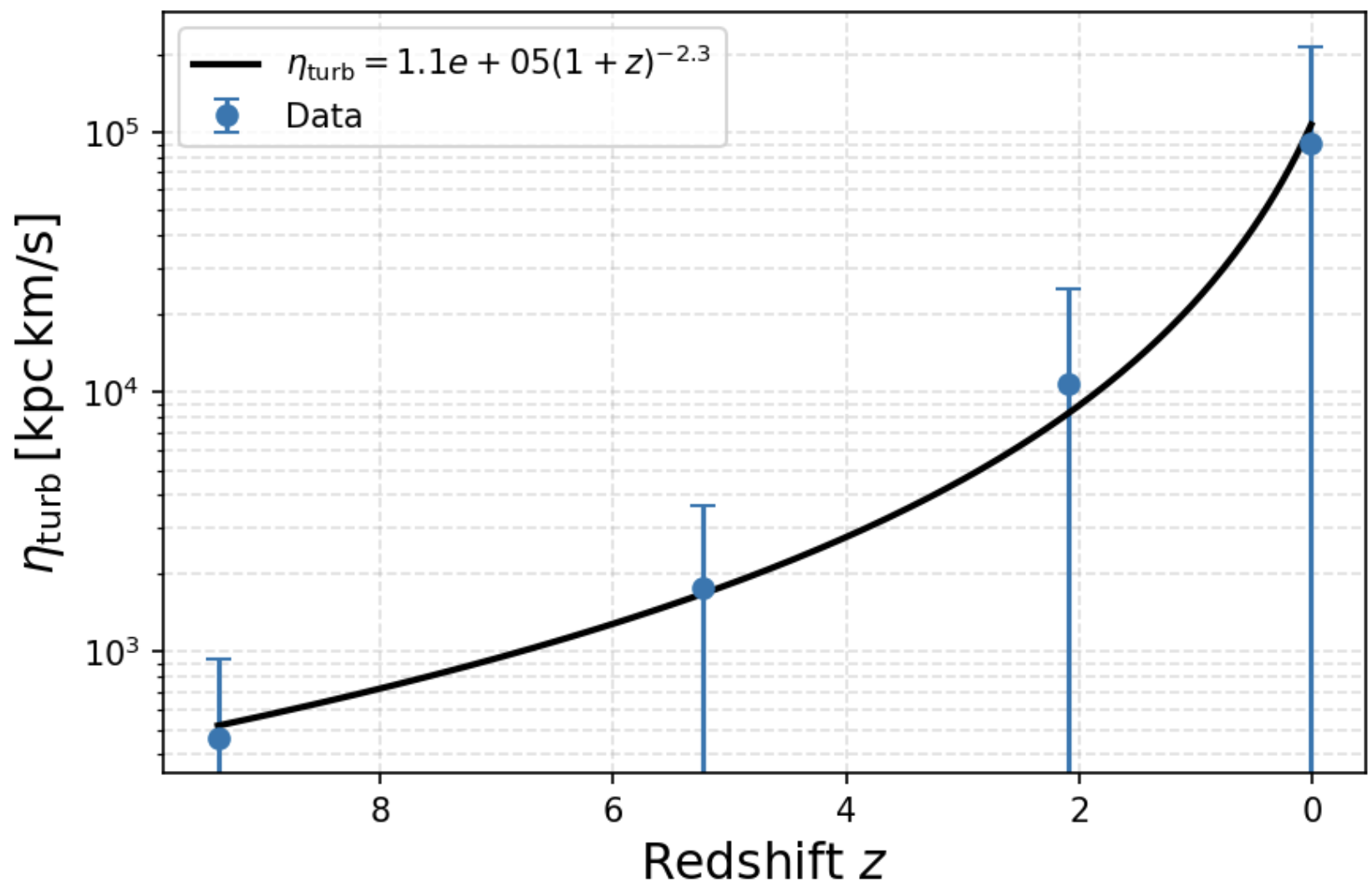}

\captionsetup{
    justification=justified,
    singlelinecheck=false
}
\caption{\justifying
Best-fit relation for the turbulent magnetic diffusivity, $\eta_{\rm turb}$, as a function of redshift. The solid curve shows the least-squares fit given by Eq.~\eqref{eq:eta_turb}, while the data points show the values estimated from the TNG50-1 simulation at $z=9.39$, $5.23$, $2.10$, and $0.01$. The error bars indicate the estimated dispersion of the turbulent magnetic diffusivity.
}
\label{fig:mu_eta_grid}
\end{figure*}

Taking the estimated dispersions into account, we perform a least-squares fit to the data. Figure~\ref{fig:mu_eta_grid} shows the resulting best-fit relation,
\begin{equation}
    \eta_{\rm turb}(z)
    =
    1.1\times10^{5}
    (1+z)^{-2.3}
    \,[\mathrm{kpc\,km\,s^{-1}}].
\label{eq:eta_turb}
\end{equation}
In the next section, we use this to evaluate the redshift evolution of the characteristic scale of magnetic fields diffused from galaxies.

\begin{table*}

\captionsetup{
    justification=justified,
    singlelinecheck=false
}
\caption{\justifying
Turbulent diffusivity derived from the TNG50-1 simulation.
The table lists the turbulent velocity $u_{\rm turb}$, the turbulent driving scale $\lambda_{\rm turb}$, and the corresponding turbulent magnetic diffusivity $\eta_{\rm turb}$ at each redshift.
}
\label{tab:turbulent_diffusivity}

\begin{ruledtabular}
\begin{tabular}{lccc}
      & $u_{\rm turb}\,[\mathrm{km\,s^{-1}}]$
      & $\lambda_{\rm turb}\,[10^3\,\mathrm{kpc}]$
      & $\eta_{\rm turb}\,[10^2\,\mathrm{kpc\,km\,s^{-1}}]$\\
      \hline
      $z=9.39$
        & $1.82 \pm 0.95$
        & $0.761 \pm 0.661$
        & $4.62 \pm 4.68$\\

        $z=5.23$
        & $3.64 \pm 2.67$
        & $1.45 \pm 1.16$
        & $17.6 \pm 19.0$\\

        $z=2.10$
        & $9.31 \pm 10.3$
        & $3.49 \pm 2.52$
        & $108 \pm 143$\\
      
      $z=0.01$
        & $20.7 \pm 25.4$
        & $13.1 \pm 7.95$
        & $903 \pm 1237$\\
      
    \end{tabular}%
\end{ruledtabular}
\end{table*}

\section{Propagation of Galactic Magnetic Fields into Voids}

In this section, we evaluate the redshift evolution of magnetic fields diffusing from galaxies using the turbulent magnetic diffusivity derived in Sec.~III. Substituting Eq.~\eqref{eq:eta_turb} into Eq.~\eqref{eq:screening_wavenumber}, we first calculate the screening scale $k_{\rm s}$ and the corresponding screening length $l_{\rm s}=2\pi/k_{\rm s}$. We then evaluate the resulting magnetic-field profile given by Eq.~\eqref{eq:B_screened_final} and assess the efficiency of magnetic-field transport into cosmic voids.

For the radius of the initially magnetized galactic region, $r_{\rm g}$, we take the empirical relation for the galaxy size proposed by Ref.~\cite{shibuya2015morphologies}. The redshift evolution of the comoving galaxy radius is given by
\begin{equation}
r^{\rm com}(z)
=
r_0(1+z)^{-0.1}.
\end{equation}
We set $r_0=10\,{\rm kpc}$ and adopt
$r_{\rm g}=r^{\rm com}(z_{\rm ini})$
for the radius of the initially magnetized galactic region.
We furthermore assume an initial magnetic-field strength of
$B_0=1\,\mu{\rm G}$.

\begin{figure*}[t!]
\centering

\includegraphics[width=.55\textwidth]{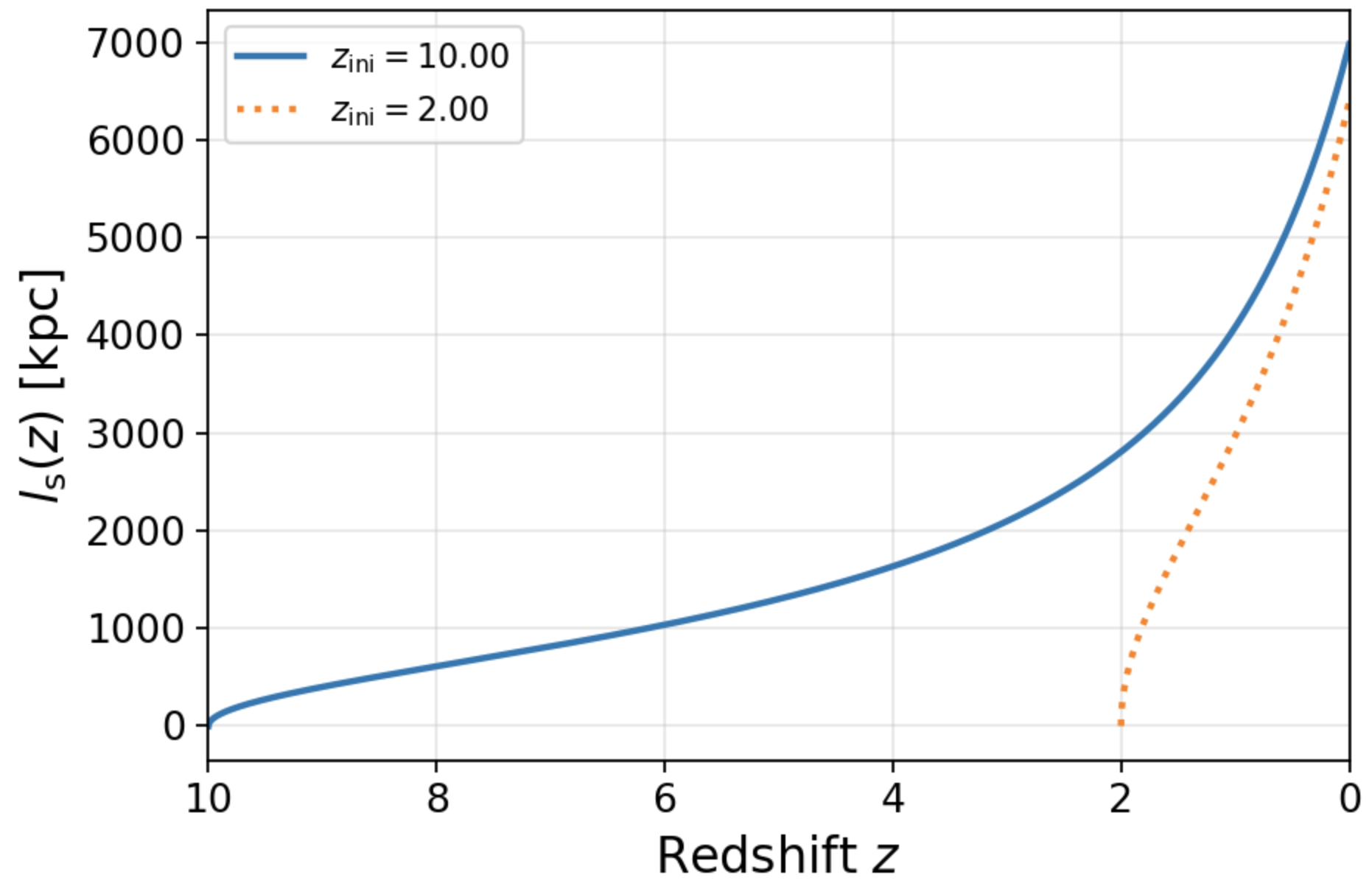}

\captionsetup{
    justification=justified,
    singlelinecheck=false
}
\caption{\justifying
Magnetic screening length $l_{\rm s}$ as a function of redshift. The solid and dotted curves correspond to $z_{\rm ini}=10$ and $z_{\rm ini}=2$, respectively. In both cases, $l_{\rm s}$ increases toward lower redshifts owing to the growth of the turbulent magnetic diffusivity.
}
\label{fig:l_s}
\end{figure*}

As shown in Fig.~\ref{fig:l_s}, the magnetic screening length grows monotonically toward lower redshifts.
This reflects the increase of the turbulent magnetic diffusivity during cosmic structure formation.
At the present epoch, the screening length reaches
$l_{\rm s}\simeq 7\times 10^3\,{\rm kpc}$
for $z_{\rm ini}=10$
and
$l_{\rm s}\simeq 6.4\times 10^3\,{\rm kpc}$
for $z_{\rm ini}=2$.

The weak difference between the two initial-redshift cases indicates that the present-day screening length depends only weakly on the initial redshift. This suggests that the resulting magnetic-field distribution is relatively insensitive to the uncertain epoch of galactic magnetic-field formation. 
The characteristic diffusion scale of galactic magnetic fields is determined primarily by the late-time evolution of the turbulent magnetic diffusivity.

\begin{figure*}[t!]
\centering

\includegraphics[width=.48\textwidth]{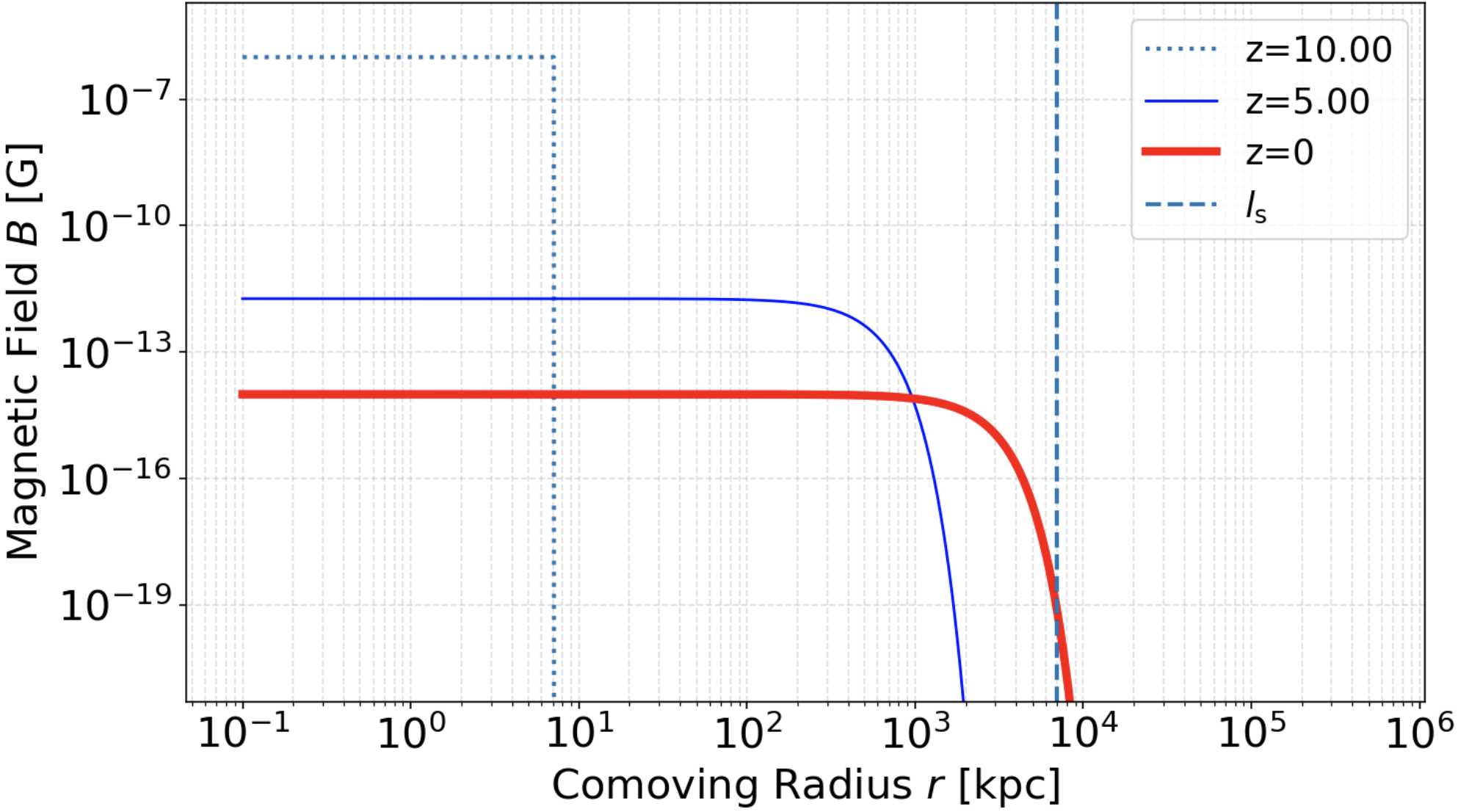}
\includegraphics[width=.48\textwidth]{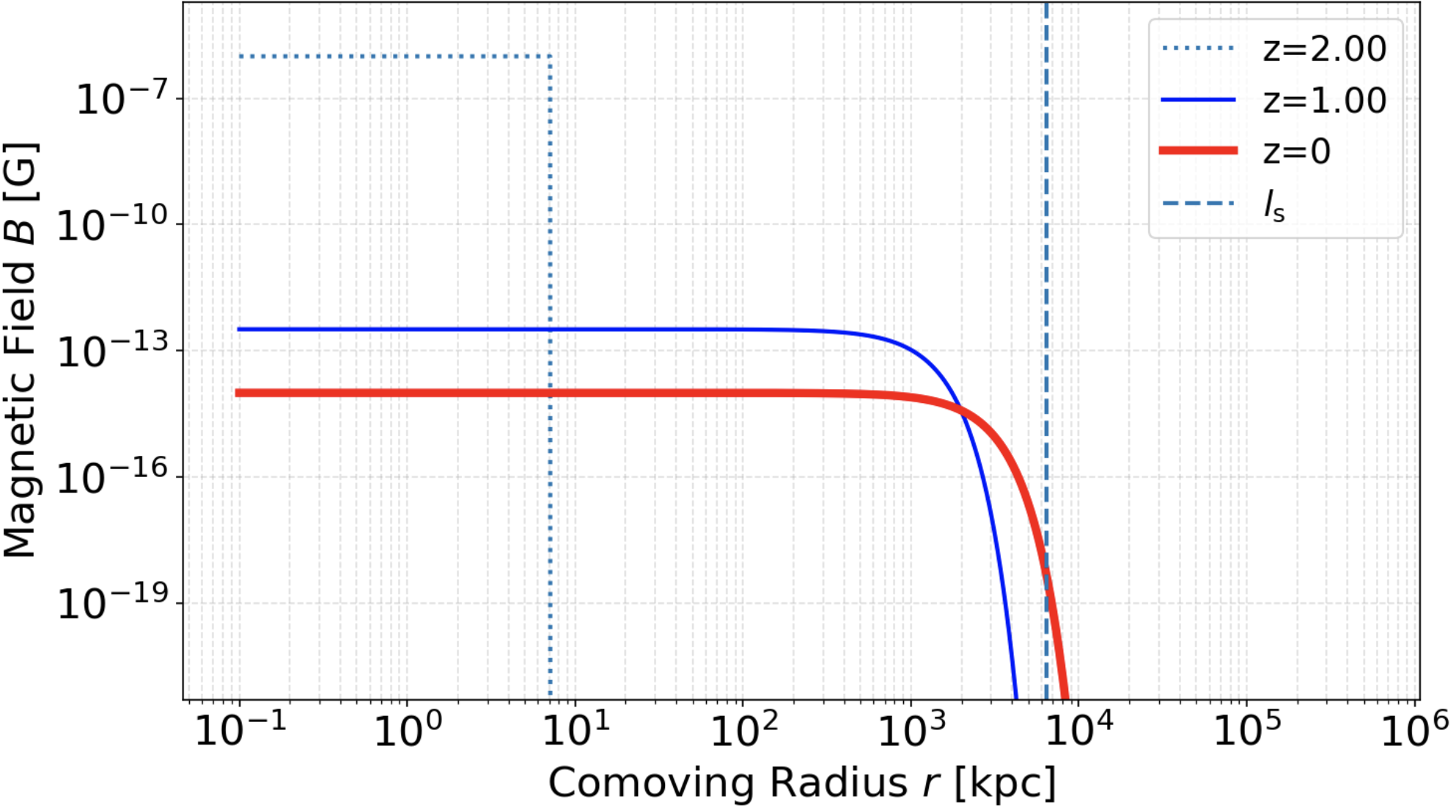}

\captionsetup{
    justification=justified,
    singlelinecheck=false
}
\caption{\justifying
Radial dependence of the physical magnetic-field strength at different redshifts. The left panel shows the evolution from $z_{\rm ini}=10$ to $z=0$, while the right panel shows the evolution from $z_{\rm ini}=2$ to $z=0$. The dashed curves represent the initial magnetic-field profiles and the solid curves the evolved profiles. The vertical lines indicate the present-day magnetic screening length $l_{\rm s}$.
}
\label{fig:radial_profile}
\end{figure*}

Using the screening wavenumber obtained from Eq.~\eqref{eq:eta_turb}, we evaluate the magnetic-field profile in Eq.~\eqref{eq:B_screened_final}.
Figure.~\ref{fig:radial_profile} shows the radial dependence of the physical magnetic-field strength at different redshifts. As the Universe evolves, the magnetic field gradually spreads beyond the initially magnetized galactic region. However, the field strength decreases rapidly with distance from the galaxy.

The vertical lines indicate the present-day magnetic screening length, $l_{\rm s}$. In both the $z_{\rm ini}=10$ and $z_{\rm ini}=2$ cases, the magnetic field is already suppressed to approximately $10\%$ of its initial value at $r\simeq l_{\rm s}/2$, and to about $10^{-4}$ at $r\simeq l_{\rm s}$. This demonstrates that $l_{\rm s}$ provides a useful indicator of the effective propagation scale of galactic magnetic fields.

Furthermore, the damping behavior is remarkably similar for the two initial-redshift cases. Consistent with the weak redshift dependence of $l_{\rm s}$ discussed above, the resulting magnetic-field profiles are nearly independent of the assumed epoch of magnetic-field generation.

\section{Summary and Discussion}

In this paper, we investigated whether galactic magnetic fields can be transported into cosmic voids through turbulent magnetic diffusion. To address this question, we derived an analytical solution of the induction equation, which relates the spatial evolution of magnetic-field diffusion to the turbulent magnetic diffusivity.

We then estimated the turbulent magnetic diffusivity from the IllustrisTNG simulation by evaluating the characteristic turbulent velocity and turbulent scale of the intergalactic medium. The resulting diffusivity increases toward lower redshifts and is well described by a power-law function of redshift.

Using the obtained turbulent diffusivity, we evaluated the resulting distribution of magnetic fields diffusing from galaxies. We found that the magnetic screening length provides a useful measure of the effective propagation scale of galactic magnetic fields. In particular, the magnetic field strength decreases to approximately $10\%$ of the initial galactic magnetic-field strength at a distance corresponding to half of the screening scale, and to about $10^{-4}$ at the screening scale itself.

Applying the turbulent diffusivity obtained from the IllustrisTNG simulation, we concluded that 
the resulting present-day magnetic screening length reaches
$l_{\rm s}\simeq 6$--$7\,{\rm Mpc}$.
This corresponds to a significant fraction of the characteristic size of cosmic voids and is 
roughly twenty times larger than previous estimates based on a constant turbulent diffusivity. Our results therefore suggest that turbulent diffusion can play a more important role in the magnetization of cosmic voids than has been inferred from earlier diffusion estimates.

Our model considers diffusion from a single galaxy and assumes a fixed initial magnetic field. In reality, multiple galaxies surrounding a void may contribute to its magnetization, while galactic dynamos continuously replenish magnetic fields over cosmic time. Both effects would tend to enhance the magnetization level relative to our estimates. Therefore, the results presented here should be regarded as conservative lower limits on the contribution of galactic magnetic fields to cosmic-void magnetization. A more realistic investigation will require modeling the spatial distribution of galaxies around voids and evaluating the cumulative contribution from multiple magnetic-field sources.

We have assumed that the turbulent velocity field is not significantly modified by the magnetic field generated by the diffusion process. To assess the importance of this effect, we compare the Alfvén velocity of the diffused magnetic field with the turbulent velocity. For a representative diffused magnetic field of $B\simeq10^{-13}$ G, we find that the ratio of the Alfvén velocity to the turbulent velocity is only $10^{-5}$--$10^{-4}$ over the redshift range considered. Thus, the Alfvén velocity is much smaller than the turbulent velocity, and the magnetic field is not expected to significantly affect the turbulent motions or the turbulent magnetic diffusivity adopted in this work.

In addition to turbulent magnetic diffusion, advection associated with galactic winds and large-scale outward plasma motions may also contribute to the transport of magnetic fields into the IGM. Previous studies have shown that feedback-driven magnetized outflows can transport magnetic fields from galaxies into the intergalactic medium \cite{aramburo2021magnetization}. Such outflows exist in  the IllustrisTNG, 
but are not explicitly included in our magnetic-field evolution model, which focuses on turbulent diffusion. Therefore, advection is not necessarily less important than turbulent diffusion and may under some conditions be comparable. The combined effects of coherent advection and turbulent diffusion will be addressed in a subsequent study.

Our results highlight the importance of accurately determining the effective turbulent magnetic diffusivity. Because the magnetic screening scale depends sensitively on $\eta_{\rm turb}$, uncertainties in the turbulent diffusivity directly translate into uncertainties in the estimated extent of magnetic-field propagation into cosmic voids.

In this work, $\eta_{\rm turb}$ was obtained from the IllustrisTNG simulation. Recent studies have shown that magnetized outflows in the IllustrisTNG framework can make a significant contribution to extragalactic Faraday rotation~\cite{2023MNRAS.519.4030A}. However, comparisons with LoTSS residual rotation-measure observations suggest that the IllustrisTNG feedback model may overestimate the magnetic contamination in the intergalactic medium~\cite{2024A&A...691A..34B}. These results indicate that the effective turbulent magnetic diffusivity remains subject to important uncertainties. Therefore, in order to establish a more reliable assessment of the role of turbulent diffusion in the magnetization of cosmic voids,
we aim to calibrate $\eta_{\rm turb}$ using both cosmological MHD simulations and observational constraints on magnetized outflows and intergalactic magnetic fields in future work.

\begin{acknowledgments}
We are grateful to Shohei Saga, Kohei Kamada, and Fumio Uchida for very valuable discussions. This work is supported in part by the JSPS grant numbers 21H04467, 24K00625, and JST FOREST Program JPMJFR20352935 (KI). YY is supported by the "THERS Make New Standards Program for the Next Generation Researchers".
\end{acknowledgments}

\end{document}